\documentclass{aa}
\usepackage{graphics}
\usepackage{times}
\usepackage{mathptm}
\fontfamily{ptm}
\selectfont

\begin{document}

\thesaurus{05(10.07.2; 10.11.1; 10.07.03 \object{M~92})}
\offprints{V. Testa, testa@coma.mporzio.astro.it}

\title{Use of DPOSS data to study globular cluster halos: an
application to \object{M~92}}

\author{V. Testa \inst{1}\and S. R. Zaggia \inst{2,3} \and S. Andreon \inst{2}
\and G. Longo \inst{2} \and R. Scaramella \inst{1} \and 
S.G. Djorgovski \inst{4} \and R. de Carvalho \inst{5}}

\institute{  Osservatorio Astronomico di Roma, via Frascati, 33
	I-00040 Monteporzio Catone, Italy
   \and Osservatorio Astronomico di Capodimonte, via Moiariello 16,
	I-80131 Napoli, Italy
   \and European Southern Observatory, K. Schwarzschild Str. 2, 
	D-85748, Garching, Germany           
   \and Department of Astronomy, California Institute of Technology, 
	MS 105-24, Pasadena, CA 91125, USA
   \and Observatorio Nacional de Rio de Janeiro, Rio de Janeiro,Brazil}

\date{Received 17 Aug 1999/ Accepted 25 Jan 2000}

\titlerunning{Extra-tidal halos in GCs: \object{M~92}}
\authorrunning{V.~Testa et al.}

\maketitle

\begin{abstract} We exploited the large areal  coverage offered by the
Digitized Palomar Observatory Sky Survey to analyze  the outermost  regions 
of  the  galactic
globular cluster   \object{M~92} (\object{NGC~6341}).  Two independent
photometric reduction   programs (SKICAT  and  DAOPHOT)  were  used to
construct  a  color-magnitude  diagram and a surface   density  
profile for this
cluster, based on  J- and F-band DPOSS  plates. A strong similarity  has 
been found in  the performance of the  two  programs in  the low--crowded
outermost cluster regions.  After removing the background contribution,
we  obtained the cluster  outer surface density profile
down    to a  surface brightness  magnitude    of 
$\mu_{\mathrm V} \sim$  31 mag arcsec$^{-2}$ and matched it with the 
inner profile  of Trager et al. (\cite{Tra95}). The profile shapes match 
very well:  since our data are uncalibrated, the  shift in magnitudes 
between  the profiles  has been also used to calibrate our profile. The 
analysis shows that the cluster has
an extra tidal halo  extending out to $\sim 30\arcmin$  from the  cluster
center at a $3~\sigma$ level over the background  noise. This halo is revealed
to be almost circular.

\keywords{globular clusters: general - globular clusters: individual: M~92 - 
Galaxy: kinematics and dynamics}
\end{abstract}

\hbadness 10000


\section{Introduction}

The tidal radii  of  globular clusters  (GCs)  are
important tools for understanding the complex interactions of GCs with
the Galaxy.  In fact, they have traditionally been  used to study the
mass distribution of the galactic halo (Innanen et al. \cite{Inn83}), 
or to deduce GCs
orbital parameters (Freeman \& Norris \cite{Fre81}; Djorgovski et al.
 \cite{DJ96}).  Tidal  radii have usually been
{\it   estimated} (only in   few  cases {\it  directly} measured),  by
fitting  King models to    cluster density profiles   rarely
measured from  the inner regions out to  the tidal radius,  because of
the nature of the photographic material, that  prevented any measure in
the cluster center, and the small  format of the first digital cameras.
 Only  in the last  few years, the advent of
deep digitized sky  surveys and wide  field digital detectors has allowed
us to  deal with   the overwhelming   problem  of
contamination from field stars and  to probe the outer region
of GCs directly (Grillmair et al. \cite{Gri95}, hereafter G95; 
Zaggia et al. \cite{Zag95}; Zaggia et al. \cite{Zag97}; Lehman \& 
Scholz \cite{LS97}).   The study of tidal  tails in
galactic satellites is gaining interest  for many applications related
to the derivation    of  the galactic  structure and   potential,  the
formation and evolution of the galactic halo, as well as the dynamical
evolution  of the clusters  themselves.   Recent determinations of 
proper motion for some  globular clusters  with HIPPARCOS have  made it
possible to  estimate   the orbital parameters  of  a  good  number of
them (Dinescu et al. \cite{Din99}).  This  helps to clarify  the
nature and  structure   of tidal  extensions  in GCs.   

In  principle,
available  tools  to enhance cluster star   counts against field stars
rely   on the color-magnitude  diagram   (CMD), proper motions, radial
velocities, or a combination of the three techniques.  The application
of these techniques to  GCs have led   to the discovery that tidal  or
extra-tidal  material  is a common  feature: Grillmair (\cite{Gri98}), for
instance,  reported the discovery   of  tidal tails in   16 out  of 21
globular clusters.  Interestingly, signature  of the presence of tidal
tails in GCs has also been found in four GC's  in M31  (Grillmair 
\cite{Gri96}).  For galactic clusters, the discovery was made by using 
a selection in the  CMD of  cluster stars   on catalogs  extracted from  
digitized photographic datasets.  The CMD selection  technique is  an  
economical and powerful method to detect  GC  tails, since it significantly  
decreases the number of background and/or foreground objects.

\begin{figure*}[t]
\resizebox{10cm}{!}{\includegraphics{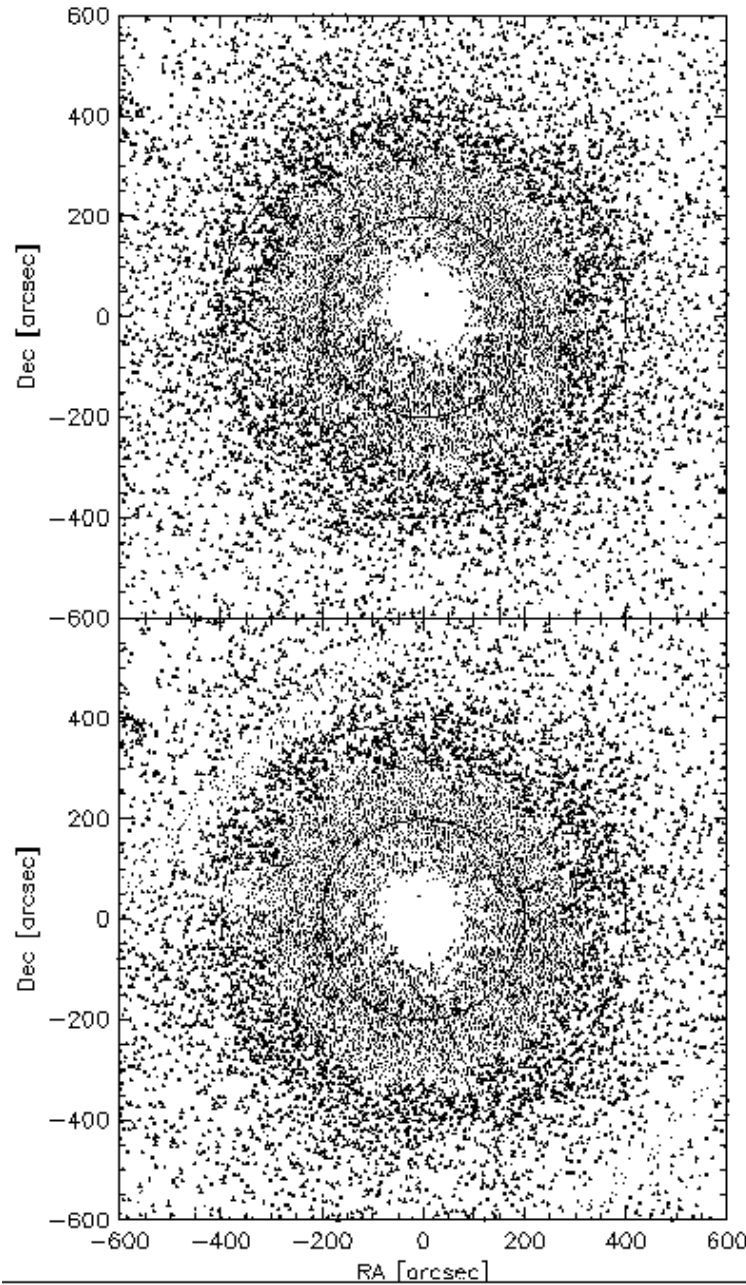}} 
\parbox[b]{55mm}{
\caption{A  comparison  of objects   detected in the   inner parts  of
\object{M~92} by SKICAT (filled  triangles) and DAOPHOT (dots) for the
two  different plates: $J$  ({\it upper  panel})  and $F$ ({\it  lower
panel}).   The inner  circle (continuous) marks  the circular aperture
where plate detections cannot be used.    The outer circle  (short-dash)
marks  the annular region where  crowding correction  is important. In
both panels  North is up and  East is to  the right.  The two diagonal
bands  in the lower  panel indicate the  satellite tracks where SKICAT
detects no objects.}
\label{fig1}}
\end{figure*}

In order  to test the  feasibility of a  survey of most GCs present in
the Northern hemisphere, we applied  the CMD technique to the galactic
globular  cluster  \object{M~92} (NGC~6341), with the aim of 
measuring  the tidal radius  and searching for  the possible presence  of
extra-tidal  material.   We used   plates  from the  Digitized  Second
Palomar Sky Survey (hereafter DPOSS), in the framework of the CRoNaRio
(Caltech-Roma-Napoli-Rio de Janeiro) collaboration
(Djorgovski et al. \cite{DJ97}, Andreon et al. \cite{And97}, 
Djorgovski et al. \cite{DJ99}). A previous account on this work was
given in  Zaggia et al. (\cite{Zag98}).  This is the first of a series  
of  papers dedicated to the subject --an  ideal application 
for this kind of all-sky surveys.

\begin{figure*}[t]
\resizebox{10cm}{!}{\includegraphics{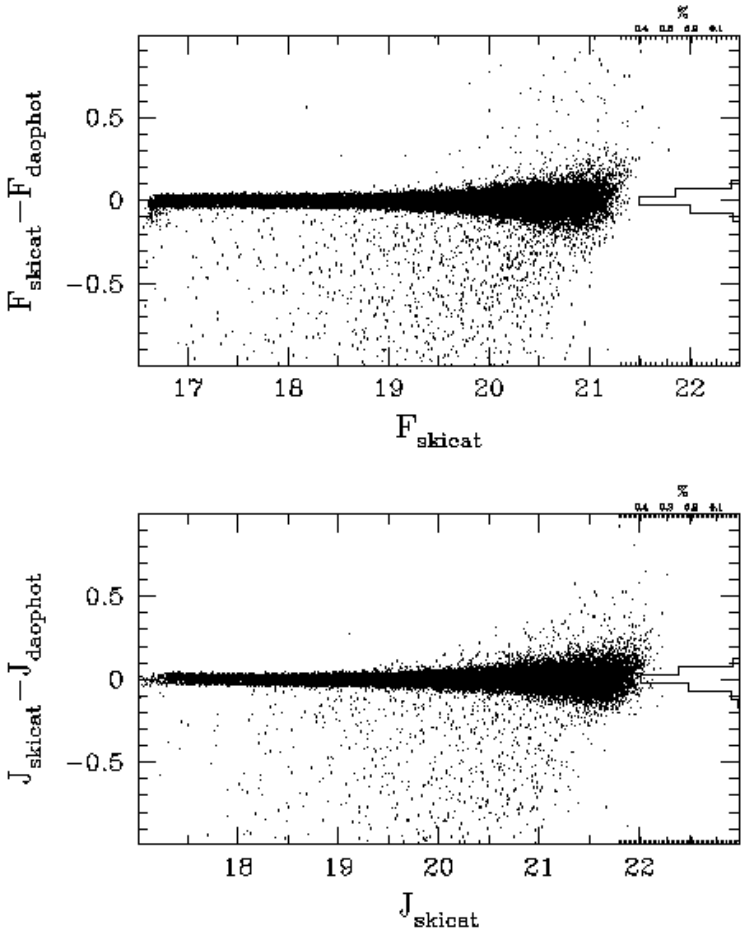}} \hfill 
\parbox[b]{55mm}{
\caption{Comparison between SKICAT   $M_{Core}$ and DAOPHOT   aperture
magnitude in the filters $F$ ({\it upper  panel}) and $J$ ({\it lower
panel}).The histograms of the distributions, expressed in percentage of the
total, are reported at the right edge of the plots.}
\label{fig2}}
\end{figure*}

\section{The Color$-$Magnitude Diagram}

The material used in this work are the $J$ and $F$ DPOSS plates of the
field 278. For each band, we extracted from the whole digitized plate
a sub-image (size: $8032 \times 8032$ pixels), corresponding to an area
of $136\arcmin \times 136\arcmin$ , with  a pixel size of $1\arcsec$,
centered on \object{M~92} at coordinates (Harris \cite{Har96}):

\[
\alpha _{J2000}=17^{h}\;17^{m}\;07.3^{s}
\]
\[
\delta _{J2000}=+43^{o}\;08{\arcmin }\;11.5{\arcsec}
\]

\begin{figure*}[t]
\resizebox{12cm}{!}{\includegraphics{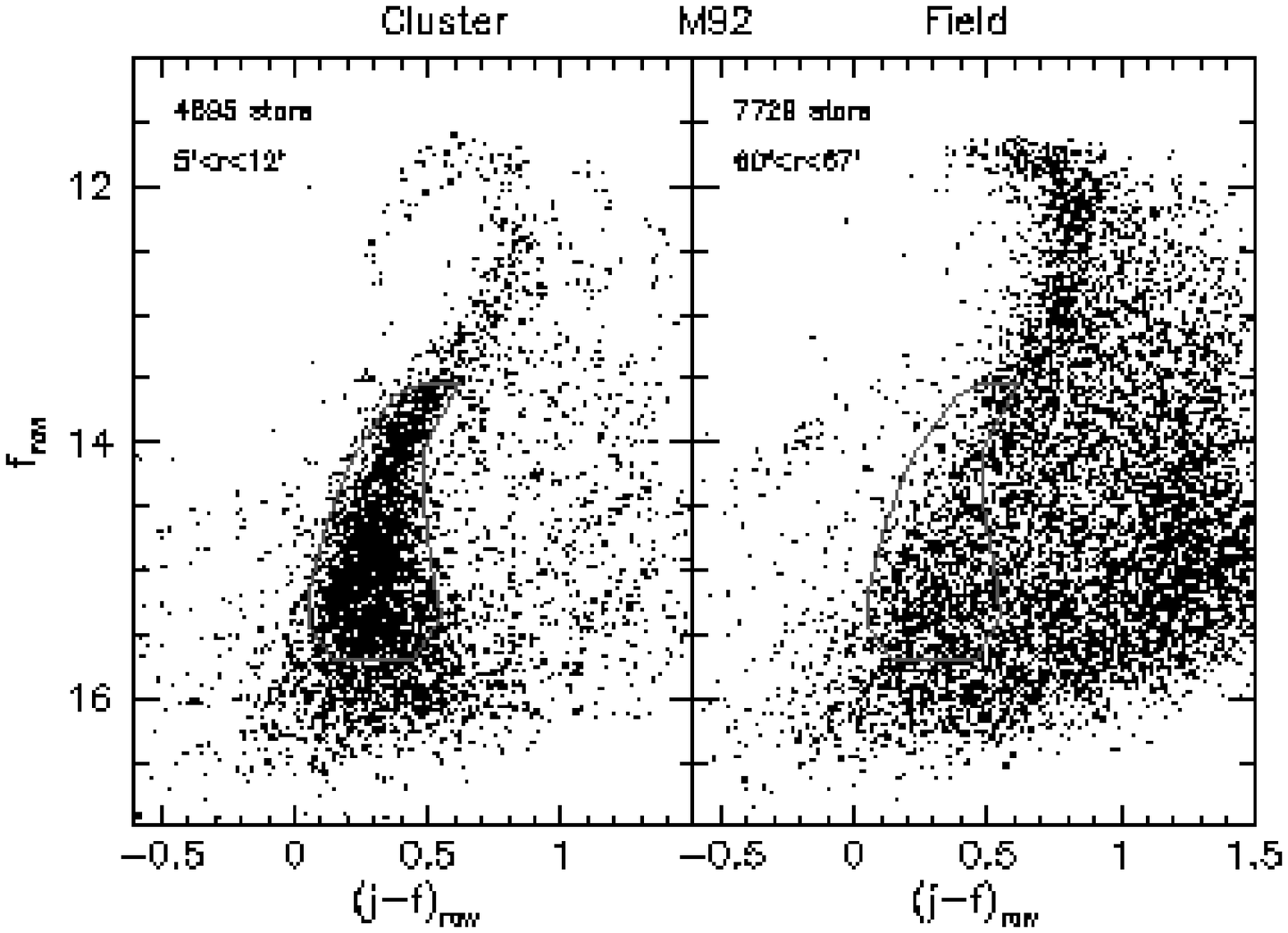}} \hfill 
\parbox[b]{55mm}{
\caption{Color magnitude diagram of the stars detected in the field of
the globular cluster \object{M~92}.  {\it Left panel:}  CMD of \object{M~92} 
in an annular region at $5' < r < 12'$.  {\it Right panel:}  background field 
CMD in an annulus at $60' < r < 67'$.}
\label{fig3}}
\end{figure*}

The  two images  were linearized by using  a density-to-intensity (DtoI)
calibration curve, provided by the sensitometric spots available on the
DPOSS plates.    The  $F$ plate  is contaminated  by  two very similar
satellite tracks (as an  alternative, the  two  tracks come from  a
high  altitude  civil   airplane)   lying $\sim  9\arcmin$   and $\sim
13\arcmin$  from the cluster   center  and crossing   the field in   a
South-East/North-West direction.  The effect  of  these tracks can  be
seen as  empty strips on  the lower  panel  of Fig.~\ref{fig1}.  Other
thin, fainter tracks  and some galaxies  are present  on the same
plate, but at   larger distances  from  the cluster  core   region.  We
applied  the  CMD   technique  to  datasets obtained   with  different
astronomical   packages, in order  to  test  the reliability of  object
detection and photometry   in  crowded stellar fields.   On  the DPOSS
plates containing \object{M~92}, we  used both the SKICAT  and
DAOPHOT packages.  SKICAT,     written    at  Caltech  (see Weir et al. 
\cite{Wei95a}, and refs. therein), is the standard software
used   by the CRoNaRio collaboration for the DPOSS plate processing and 
catalog construction.  DAOPHOT is a well-tested program for stellar photometry,
developed by Stetson (\cite{Ste87}), and widely used by stellar astronomers. 
In this work  we have  used DAOPHOT only to obtain  aperture photometry,
with APPHOT,  of objects  detected with  the DAOFIND  algorithm on the
DPOSS plates.

\subsection{The data set}

The SKICAT output catalog only contains objects classified as
stars   in both      filters.    For    each object,     we      used
$M_{\mathrm{Core}}$  (the magnitude  computed from the  central
nine pixels), because the other aperture  magnitude is measured on an area
far too large  for crowded regions. The final  SKICAT catalog  consists of
108779 objects.   Since SKICAT is optimized for the detection of faint
galaxies, in the  present case we needed to  test its  performances in
crowded stellar fields to ensure that it properly detected the stellar
population around the cluster.

Thus, SKICAT has been compared to DAOPHOT, which is
specifically designed for crowded fields stellar photometry and has
been repeatedly tested in a variety of environments, including
globular clusters.  The DAOPHOT dataset was built using aperture
photometry on the objects detected with the DAOFIND. The threshold was
set at $3.5~\sigma$, similar to the one used by SKICAT.  Aperture
photometry was preferred to PSF fitting photometry, due to the large
variability of the DPOSS point-spread function which makes the
PSF photometry less accurate than the aperture photometry.  We used an
aperture of $1.69$ pixels of radius, corresponding to an area of
approximately 9 pixels, i.e.  equivalent to the area used by
FOCAS/SKICAT to compute $M_{\mathrm{Core}}$.  Indeed, the
advantages of using PSF fitting are more evident in the central and
more crowded regions of the cluster, while we are mainly interested in
the outskirts, where crowding is less dramatic. Thus, we adopted the
results from the aperture photometry, and we refer to this dataset as
the DAOFIND+PHOT dataset.

The  total number of objects   detected in the  $J$ and  $F$ plates is,
respectively, 240138 and 253977. The larger number of objects detected
by DAOFIND, compared  to those from SKICAT, is mainly  due to the better
capacity of DAOFIND in detecting objects in  the crowded regions of the
core. In the case of DAOFIND, since the convolution kernel, which is set 
essentially by the pixel size and seeing value, is much smaller than
in SKICAT, we also have objects measured near the satellite tracks.

\begin{figure*}[t]
\resizebox{12cm}{!}{\includegraphics{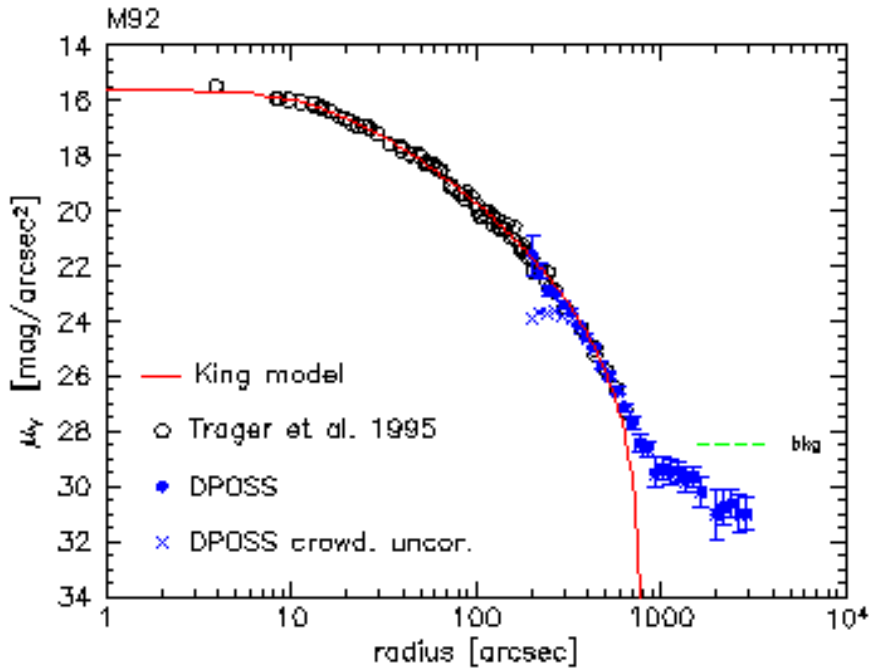}} \hfill 
\parbox[b]{55mm}{
\caption{\object{M~92} radial density profile. Open dots, 
Trager et al. (1995) data; filled dots, DPOSS SCs; crosses, crowding 
uncorrected SCs; solid line, isotropic King model.}
\label{fig4}}
\end{figure*}

The FOCAS/SKICAT  and  DAOFIND+PHOT   aperture photometry   were  then
compared, and the results are  shown in Fig.~\ref{fig2} where the SKICAT
aperture magnitude is plotted versus the difference between itself and
$M_{\mathrm{Core}}$.  The  average difference is zero, with an
error distribution   typical of this kind  of  tests,  i.e.  a
fan-like  shape with growing  dispersion  at  fainter magnitudes.  The
distribution of F magnitudes in Fig.~\ref{fig2} clearly shows the effects of
saturation at   the bright end,  and in  both plots  there are several
outliers, owing to the field crowding.  In fact, these outliers
are much more  concentrated  in the inner  $12 \arcmin$,  where  their
density   is 0.439 arcmin$^{-2}$, than   at  larger distances  from the
center, where the density drops to 0.022 arcmin$^{-2}$. 
These  objects  are mostly classified as non  stellar  by SKICAT and DAOPHOT,
since they are either foreground galaxies or, more often,  unresolved
multiple objects, and were rejected in the final catalogs.  Their area
is taken into account later on, when we compute the  effective area of the
annuli in the construction of the radial profile. The outliers show an
asymmetric distribution with   SKICAT  magnitudes being  brighter   at
bright   magnitudes   and viceversa at     fainter magnitudes. This is due
to two reasons: in the case of large objects, SKICAT splits them into
multiple entries, but keeps the $M_{\mathrm{Core}}$ value of 
the originally detected (big)  object; at fainter magnitudes,
where objects are small, $M_{\mathrm{Core}}$ is computed on a
number  of pixels less  than 9, while the  aperture  photometry of the
objects in the DAOFIND catalog are always computed on a circle of 1.69
pixel radius.However, the
contribution of these outliers to the counts is far below 1 percent
of the total, as can be seen from the histogram plotted on the right hand
side of fig. \ref{fig2}.

The  above analysis shows that SKICAT  catalogs are, after a suitable
cleaning, usable   ``as they  are''  also   for studies  of moderately
crowded  stellar  fields. We shall  use  the
DPOSS-DAOPHOT dataset  because it can  better detect 
objects in  highly crowded fields, which allows  us  to probe into the
inner ($200 \arcsec < r < 400 \arcsec$) regions of  the cluster and 
merge our star counts profile with the published one of Trager et al.
(\cite{Tra95}, hereafter T95).

\subsection{The color-magnitude diagram}

In order to build the CMD of the cluster, individual catalogs were
matched by adopting a searching radius of 5 $\arcsec$, and keeping 
the matched object with the smallest distance.  The derived CMD 
is shown in Fig.~\ref{fig3} for two annular
regions: the inner one, between 5$\arcmin$ and 12$\arcmin$ from the center 
(left-hand  panel) and the outer one, referring to the background, 
between 60$\arcmin$ and 67$\arcmin$ (right-hand panel). The \object{M~92}
turn-off region, as well as  part of the horizontal branch, are clearly
visible. At bright magnitudes, the giant branch turns to the blue, due to 
plate saturation.  At large angular distances
from the center of \object{M~92}, most objects are galactic stars with
only a small contribution from the cluster.

For  reducing the background/foreground field  contribution, we used an
approach similar  to that of Grillmair et al. \cite{Gri95}.  First of all,
We selected an annular region ($200 \arcsec <r<300 \arcsec$) around the  
cluster center to find the best fiducial CMD sequence  of the  cluster
stars. Then, the CMD of  this region was compared with  the CMD of the
field at   a distance greater than  $\simeq1^\circ$  from  the cluster
center.   The two CMD's   were normalized by  their area  and then  we
binned the CMD and computed the $S/N$ ratio of each  element, just like 
in  G95 (their Eq. 2).  Finally, we then
obtained the final  contour of the best  CMD  region by  cutting at a
$S/N \simeq  1$.  Contours are shown  in Fig.~\ref{fig3},  on which a solid
line marks  the CMD region  used  to select the  ``bona fide'' cluster
stars as described above. We must say  that this CMD selection
is  not aimed at finding  all  the stars in the   cluster but only  at
the best possible enhancement of the cluster stars as compared to the field
stars. This is why the region of the sub-giant/giant branch is not
included, since here the cluster stars are fewer than in the field.  
By extracting objects at any distance
from the center  of the cluster, in the  selected CMD  region, the field
contamination is reduced by a factor of $\sim 7$. In absence of strong
color gradients, the  fraction of lost  stars  does not depend  on the
distance from the center.

\section{Extra-tidal excess in \object{M~92}}

\subsection{Radial density profile}

As first   step, we built  a  2-D star  density  map by  binning the
catalog in step of $1\arcmin \times 1\arcmin$.  Then, we fitted a polynomial
surface to the background, selecting only the outermost regions of the
studied area. The background correction is expressed in the same units
of 2-D surface density counts,  and can be direclty applied to the raw 
counts.
A tilted  plane  was sufficient  to   interpolate the  background star
counts (SCs).   Higher-order polynomials   did   not  provide   any
substantial improvement over the adopted solution. We compared the
fitted background with IRAS maps at 100$\mu$, but we did not find any
direct signs of a correlation between the two. Rather, the direction of 
the tilt is consistent with the direction of the galactic center. 
Hence, the tilt of the background, which is however very small 
($\sim 0.01 \mathrm{mag/arcmin}$), can be considered as due to the 
galactic gradient.

The cluster  radial density profile  was obtained from  the background
subtracted SC's  by counting  stars   in annuli  of  equal logarithmic
steps.  The  uncorrected surface density  profile  (hereafter SDP) is
expressed as:
$$SDP_{i}^{\mathrm{uncorr}}=-2.5*log(N_{i,i+1}/A_{i,i+1})+const,$$        
where  $N_{i,i+1}$ indicates  the  number  of  objects  in the annulus
between   $r_{i}$  and $r_{i+1}$,  and  $A_{i,i+1}$ 
 the   area of the
annulus. The constant was determined  by matching the profile with the
published profile of T95 in the overlap range.  
The effective radius at each point of the profile is given by:
$$r_{i}^{\mathrm{eff}}=\sqrt{{1 \over 2}\times(r_{i}^2+r_{i+1}^2)}.$$

The SDP must now be  corrected for crowding.
When dealing with photographic  material, it is not
possible to  apply the  widely  known artificial   star
technique used with CCD data. Therefore,  we used a procedure similar
to the ones  described in Lehman \& Scholz (\cite{LS97}) and  Garilli et al.
(\cite{Gar99}):  we estimated the
area occupied by  the objects in each  radial annulus by selecting all
the pixels brighter than the background  noise level plus three sigmas,
and considered as virtually uncrowded the external annuli in which the
percentage (very  small,  $\simeq0.5\%$) of  filled  area did not vary
with the distance. The external  region starts at $\sim 1000
\arcsec$ from  the cluster center  with a filling factor  smaller than
$\simeq 2\%$.  After   correcting the   area covered by    non-stellar
objects,   the ratio  unfilled/filled area
gives  the crowding correction. This  correction was computed at the
effective radii of the surface brightness  profile (hereafter SBP) and
smoothed using a  spline  function.  The corrected surface  brightness
profile was then computed as:
$$SBP_{i}=SBP_{i}^{\mathrm{uncorr}}+2.5\times \mathrm{log}(1.0-frac)$$
where  $frac$ is  the  crowding  correction factor  determined at  the
i$-th$ point on the profile. 
\begin{figure*}
\resizebox{12cm}{!}{\includegraphics{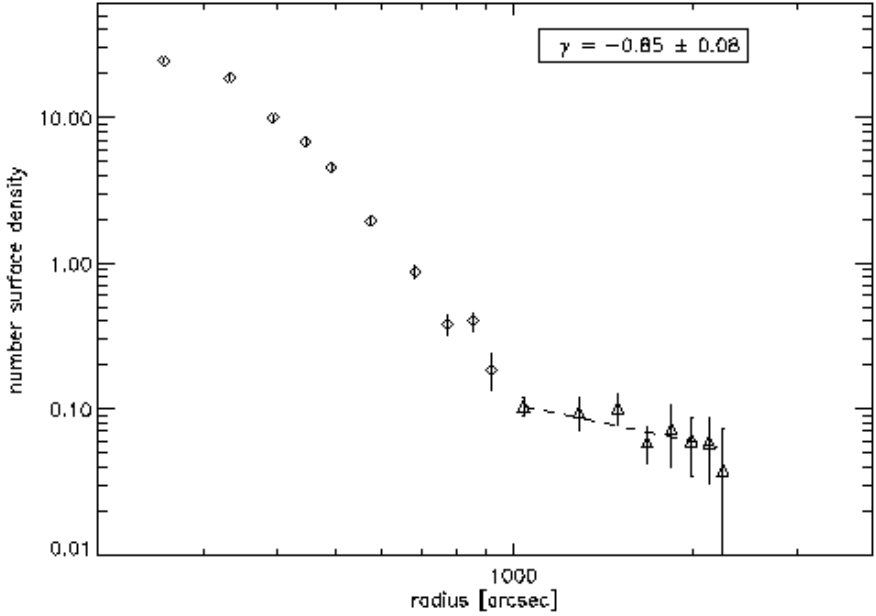}} \hfill 
\parbox[b]{55mm}{
\caption{Fit of a power law to the external  profile of \object{M~92},
expressed in number surface density to allow for an immediate comparison
with Johnston et al. (\cite{Jon99}, hereafter J99), and binned to smooth 
out oscillations due to the small dimension of the annuli.  Triangles 
indicate  the profile of the  extra-tidal  halo.  Diamonds represent 
the binned and averaged profile within the tidal radius. The dashed line 
is the fitting power law with $\gamma = -0.85 \pm 0.08$.}
\label{fig5}}
\end{figure*}

The crowding corrected SBP of \object{M~92} derived from DPOSS  data is 
shown in Fig.~ \ref{fig4}  as filled dots and the uncorrected counts as 
crosses. Table~\ref{tab1} lists the measured surface brightness profile. 
With a simple number counts normalization we joined our
profile to the one (open circles) derived  by T95,
in  order to extend the profile to the inner regions.  We then  fitted a 
single-mass King  model to our  profile. The  fitting  profile is drawn  on
Fig.~\ref{fig4}  as a continuous  line.    Our value for  the tidal
radius,  $r_{\mathrm t} =  740\arcsec$,  turned out to be   similar to the 
value
given in Brosche et al. (\cite{Bro99}), $r_{\mathrm t} = 802\arcsec$ and 
slightly smaller than the one given in  T95, 
$r_\mathrm{t} =912\arcsec$.  As it  can be seen from
the figure, DPOSS   data extend at  larger  radial distances than  the
T95 compilation and  reveal the  existence  
of a noticeable deviation  from the isotropic  King model  derived from  the
direct fitting of the SBP.  This deviation is a clear sign of the
presence of extra tidal material. We also tried fitting anisotropic 
King models to the SBP, but the fit was not as good as in the isotropic
case.

\begin{table}
\caption{Measured surface brightness profile}
\label{tab1}
\begin{tabular}{cccc}
\hline
log($r$) & $V_{SBP}^{uncorr.}$ & $V_{SBP}^{corr.}$ & $\sigma(V_{SBP})$ \\
\hline
2.307 &  24.09 &  21.84 &   0.09 \\
2.348 &  23.90 &  22.39 &   0.06 \\
2.389 &  23.96 &  23.10 &   0.09 \\
2.431 &  23.83 &  23.22 &   0.05 \\
2.472 &  24.00 &  23.62 &   0.07 \\
2.514 &  24.17 &  23.89 &   0.06 \\
2.555 &  24.50 &  24.36 &   0.06 \\
2.596 &  24.89 &  24.79 &   0.07 \\
2.638 &  25.26 &  25.19 &   0.07 \\
2.679 &  25.87 &  25.82 &   0.10 \\
2.720 &  26.25 &  26.21 &   0.09 \\
2.762 &  26.79 &  26.75 &   0.13 \\
2.803 &  27.38 &  27.34 &   0.14 \\
2.845 &  27.97 &  27.93 &   0.19 \\
2.886 &  28.70 &  28.66 &   0.30 \\
2.927 &  28.90 &  28.86 &   0.28 \\
2.969 &  29.76 &  29.72 &   0.54 \\
3.010 &  29.61 &  29.56 &   0.41 \\
3.052 &  29.75 &  29.71 &   0.41 \\
3.093 &  29.73 &  29.69 &   0.38 \\
3.134 &  30.03 &  29.98 &   0.43 \\
3.176 &  29.91 &  29.87 &   0.36 \\
3.217 &  30.46 &  30.42 &   0.51 \\
3.300 &  31.28 &  31.23 &   0.85 \\
3.341 &  31.02 &  30.97 &   0.62 \\
3.383 &  30.87 &  30.82 &   0.50 \\
3.424 &  31.26 &  31.21 &   0.63 \\
3.466 &  31.24 &  31.19 &   0.57 \\
\hline
\end{tabular}

\end{table}

At what level is this deviation significant?  The determination of the
tidal  radius of  a cluster  is still  a moot case.
While fitting a  King model  to  a  cluster density
profile,  the determination of  the  tidal  radius  comes from  a
procedure where  the   overall profile is considered,  and  internal
points weigh more than external  ones. On the one hand, this is an
advantage since the population  near the limiting radius  is a mix  of
bound stars and stars on the verge  of being stripped from the cluster
by the Galaxy tidal  potential.  On the  other hand, the tidal  radius
obtained in this way can be a poor approximation of the real one. 
In the classical picture, and in presence of negligible diffusion,  the
cluster    is  truncated  at   its  tidal    radius  at perigalacticon
(see Aguilar et al. \cite{Agu88}).  Nevertheless, Lee \& Ostriker 
(\cite{Lee87}) pointed  out that
mass loss is  not instantaneous at  the tidal radius,  and, for a given
tidal field, they expect a globular  cluster to be more populated 
than in  the corresponding King model.   Moreover, a globular cluster along
its orbit also  suffers from  dynamical shocks,  due to the crossing of the 
Galaxy disk and, in case of  eccentric orbits, to close passages near the
bulge,  giving rise to enhanced mass-loss and, later on,  to
the destruction  of  the globular cluster itself. 
Gnedin \& Ostriker (\cite{Gne97}) found that, after a gravitational shock,
the cluster expands as a whole,  as a consequence of internal  heating. 
In this  case, some stars move beyond
the   tidal  radius  but are  not    necessarily  lost, and are  still
gravitationally  bound to the  cluster.  This  could explain observed
tidal radii larger than expected for orbits with  a small value of the
perigalacticon. Brosche et al. (\cite{Bro99}) point out that the observed
limiting radii are too large to be compatible with  perigalacticon 
$r_\mathrm{t}$, and suggest
that  the appropriate  quantity  to be considered is  a  proper average of
instantaneous tidal  radii along the orbit.  It can be seen from
Fig.~\ref{fig4}   that  the     cluster  profile  deviates  from   the
superimposed King  model before the estimated  tidal radius, and has a
break in  the slope at about  $r\sim850\arcsec$, after which the slope
is constant. We shall come back later to this point. 

In  Fig.~\ref{fig5} we show  the surface density profile, expressed in 
number of stars to allow a direct comparison with J99, and binned in order to smooth  out oscillations in the  
profile, due to the small S/N ratio arising with small-sized annuli.
J99 predict that stars stripped from 
a cluster, and forming a tidal stream, show a density profile described 
by a power law with exponent $\gamma =  -1$.  We fitted a  power law of 
the type $\Sigma(r)  \propto  r^{\gamma}$ to  the  extra-tidal profile  
(dashed line).  The best fit gives  a value $\gamma  = -0.85 \pm 0.08$ 
and  is shown as a dashed  line  in Fig. \ref{fig5}.  The errors on the
profile points include also the background uncertainty, in quadrature,
so that the significance of the extra-tidal profile has been estimated in
terms of  the difference $f_i-3\sigma_i$, where $f_i$  is 
the number surface density  profile at point  $i$, and  $\sigma_i$ its
error,  which  includes  the   background  and the  signal  Poissonian
uncertainties.   This quantity is  positive for all  the points except
for the outermost one.  The fitted slope  is consistent  with  the value
proposed by J99 and in good accordance with literature values for other 
clusters (see G95 and Zaggia et al. \cite{Zag97}).

We then fit an ellipse to the extra-tidal profile in
order to derive a position angle of the tidal extension, and
checked whether the profile in that direction differs from the one
obtained along the minor axis of the fitting ellipse, to
confirm that the extra-tidal material is a tail rather than a
halo.  The best fitting ellipse, made on the ``isophote'' at the
$2~\sigma$ level from the background (approximately $1.5r_{\mathrm t}$ 
from the center of the cluster), turned out to have a very low ellipticity,
($e \leq 0.05\pm0.01$ at P.A.$ \simeq 54^\circ\pm15^\circ$).  We have also
measured the radial profiles along the major and minor axes, using an
aperture angle of $\pm 45^\circ$, in order to enhance the S/N ratio in
the counts.  The two profiles turned out to be indistinguishable
within our uncertainties. This result shows that the halo material has
a significantly different shape than the internal part of the cluster
which shows an ellipticity of $0.10\pm0.01$ at P.A. $141^\circ\pm1$ as
found by White \& Shawl (\cite{Whi87}).

\begin{figure*}[t]
\resizebox{12cm}{!}{\includegraphics{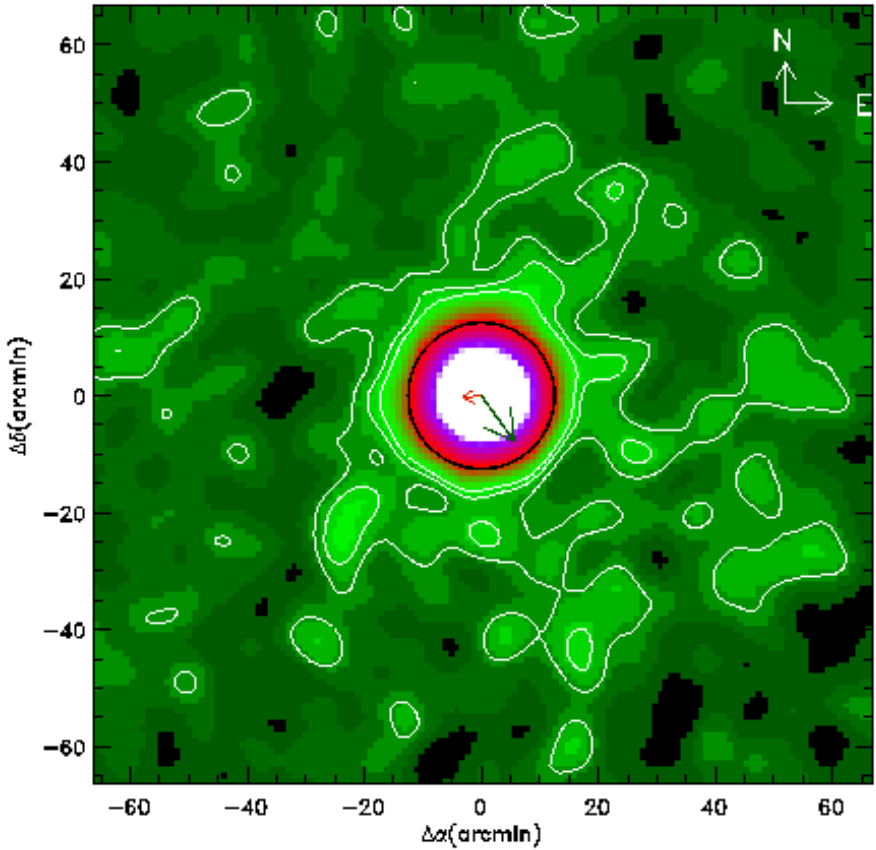}} \hfill 
\parbox[b]{55mm}{
\caption{\object{M~92} surface  density   map from  background   
subtracted star
counts. The black, thick circle is drawn at the estimated tidal radius
of   \object{M~92}.  The long, thicker  arrow indicates  the   direction of the 
galactic
center,  the thin arrow indicates the  proper motion of the cluster as
in  Dinescu et al. (\cite{Din99}).  Contours are 
  drawn at 1,2  and  3 $\sigma$ of the background.}
\label{fig6}}
\end{figure*}

\subsection{Surface density map}

In the attempt   to   shed more light  upon presence  and  
characteristics  of the extra-tidal  extension, we  used the 2-D  star
counts  map, as described at the beginning   of the previous section. We
applied a Gaussian smoothing algorithm to the map, in order to enhance
the low spatial frequencies  and  cut out  the high frequency  spatial
variations, which contribute strongly to the  noise.   We smoothed
the map using a Gaussian kernel of $6 \arcmin$. The resulting smoothed
surface density map is shown in Fig.~\ref{fig6}.  Since the background
absolute level is zero, the darkest gray levels indicate negative star
counts.   In this  image,   the probable tidal tail  of  \object{M~92}
(light-gray  pixels around the cluster)  is less prominent than in the
radial  density profile: this   is because data are not
averaged  in azimuth.  On the  map we  have  drawn three ``isophotal''
contours at 1, 2 and 3$~\sigma$ over  the background.  The fitted tidal
radius is marked  as a thick  circle and the two arrows  point toward the
galactic center (long one) and  in the direction  of  the measured 
proper motion (see Dinescu et al. \cite{Din99}). The tidal halo does not 
seem to have a preferred direction. A  marginal sign of elongation is
possibly visible along a direction  almost orthogonal to that of
the galactic center.

As pointed out in the previous section, if we  build the profile along
this direction and orthogonally to it, we do not derive clear signs of
any difference in the star count profiles  in one direction or the other,
mainly because of the small  number counts. 

On the basis of these  results, we can interpret the extra-tidal
profile of \object{M~92} as follows: at radii just beyond the fitted King
profile tidal radius, the profile resembles a halo of stars --most likely 
still tied up to the cluster or in the act of being stripped
away.  As the latter process is not instantaneous, these stars will
still be orbiting near the cluster for some time. We
cannot say whether this is due to heating caused by tidal shocks,
or to ordinary evaporation: a deep CCD photometry to study the mass
function of extra-tidal stars would give some indications on this
phenomenon.  At larger radii, the 1 $\sigma$ ``isophote'' shows a barely 
apparent elongation of the profile in the direction
SW to NE, with some possible features extending approximately towards 
S and E. Although the significance is only at 1 $\sigma$ 
level, these structures are visible and might be made up by stars escaping 
the cluster and forming a stream along the orbit.  As pointed out in 
Meylan \& Heggie (\cite{Mey97}), stars escape from the cluster from the 
Lagrangian points situated on the
vector connecting the cluster with the center of the Galaxy, thus
forming a two-sided lobe, which is then twisted by the Coriolis force.
A clarifying picture of this effect is given in Fig. 3 of 
Johnston (\cite{Jon98}).

\section{Summary and conclusions}

We investigated the presence and significance of  a tidal extension of
the brightness profile of \object{M~92}. The main results of our study are:
\begin{enumerate}

\item The presence of an extra-tidal  profile extending out to
$\sim~0.5^{\circ}$ from the cluster center, at a significance level of
$3~\sigma$ out to  $r~\sim~2000~\arcsec$.  We found no strong evidence
for preferential  direction of  elongation  of the profile.  This  may
imply that we are detecting the extra-tidal halo of evaporating stars,
which will later form a tidal stream. Moreover, the tidal tail might be
compressed along the line of sight --see, for instance, Fig.~18
of G95. In fact, G95 point out that tidal tails extend over enormous 
distances ahead and behind the cluster orbit,  and the volume density  
is subject to the open-orbit analogous of Kepler's third law:  near  
apogalacticon, stars  in the tidal  tail undergo  differential  slowing-down, 
so that  the  tail converges upon the
cluster.   Actually, most models (e.g., Murali \& Dubinsky \cite{Mur99})
predict that the  extra-tidal material should  continue to follow  the
cluster orbit and  thus take  the shape  of an  elongated  tail, or  a
stream. The  stream  has been  already   revealed in dwarf  spheroidal
galaxies  of  the local  group  (Mateo et al. \cite{Mat98}), but  
whether the stream can also  be visible in   significantly smaller objects
like  globular clusters is currently a moot point.

\item  By constructing   the surface density    map and  performing  a
Gaussian smoothing,  the low-frequency features  are enhanced over the
background.  We find some marginal evidence  for a possible elongation
in the extra-tidal extention based on a visual inspection of this map.
This elongation  may be aligned  in a direction perdendicular
to  the Galactic center, although  we already know  that the significance of
this result is low; additional observations will be required to settle
the  issue.    A similar  displacement  is    described  in  Fig.~3 of
Johnston (\cite{Jon98}).

\end{enumerate}

Finally, we  want to stress the power  of the DPOSS material
in  conducting this kind of  programs, either by using the standard output
catalogs, as they come out  from the processing  pipeline, or the
specific  re-analysis of the digitized  plate scans.  In the future we
will extend this study to most of the globular clusters present on the
DPOSS plates.

\begin{acknowledgements}
SGD   acknowledges support from the   Norris Foundation.  We thank the
whole POSS-II, DPOSS, and CRoNaRio teams for their efforts.
\end{acknowledgements}


\end{document}